\documentclass[preprint2]{aastex}

\slugcomment{Accepted to ApJ}

\shorttitle{H$\alpha$ Emission Variability in LS I +61 303}
\shortauthors{McSwain et al.}

\begin{document}

\title{H$\alpha$ Emission Variability in the $\gamma$-ray Binary LS I +61 303}


\author{M.\ Virginia McSwain\altaffilmark{1}}
\affil{Department of Physics, Lehigh University, 16 Memorial Drive E, Bethlehem, PA 18015}
\email{mcswain@lehigh.edu}

\author{Erika D.\ Grundstrom\altaffilmark{1}}
\affil{Physics and Astronomy Department, Vanderbilt University, 6301 Stevens Center, Nashville, TN 37235}
\email{erika.grundstrom@vanderbilt.edu}

\author{Douglas R.\ Gies}
\affil{Center for High Angular Resolution Astronomy and Department of Physics and Astronomy, Georgia State University, Atlanta, GA 30303-4106}
\email{gies@chara.gsu.edu}


\author{Paul S.\ Ray}
\affil{Space Science Division, Naval Research Laboratory, Code 7655, 4555 Overlook Avenue SW, Washington, DC 20375}
\email{paul.ray@nrl.navy.mil}

\altaffiltext{1}{Visiting Astronomer, Kitt Peak National Observatory. KPNO is operated by AURA, Inc.\ under contract to the National Science Foundation.}


\begin{abstract}

LS~I~$+61~303$ is an exceptionally rare example of a high mass X-ray binary (HMXB) that also exhibits MeV--TeV emission, making it one of only a handful of ``$\gamma$-ray binaries''.  Here we present H$\alpha$ spectra that show strong variability during the 26.5 day orbital period and over decadal time scales.  We detect evidence of a spiral density wave in the Be circumstellar disk over part of the orbit.  The H$\alpha$ line profile also exhibits a dramatic emission burst shortly before apastron, observed as a redshifted shoulder in the line profile, as the compact source moves almost directly away from the observer.  We investigate several possible origins for this red shoulder, including an accretion disk, mass transfer stream, and a compact pulsar wind nebula that forms via a shock between the Be star's wind and the relativistic pulsar wind.  

\end{abstract}

\keywords{accretion, accretion disks -- stars: winds, outflows -- stars: emission line, Be -- stars: individual(\object{LS~I~$+61~303$})}


\section{Introduction}

LS~I~$+61~303$ is a high mass X-ray binary (HMXB) that is also a confirmed source of very high energy $\gamma$-ray emission.  The system consists of an optical star with spectral type B0 Ve and an unknown compact companion in a highly eccentric, 26.5 day orbit \citep{aragona2009, grundstrom2007, casares2005a}.  While the system has a relatively low X-ray luminosity for a HMXB, LS~I~$+61~303$ is the 15th brightest $\gamma$-ray source included in the Fermi LAT 1-year Point Source Catalogue \citep{abdo2010}.  The Be disk interacts with the compact companion, producing orbital phase modulated emission across the electromagnetic spectrum:  TeV  \citep{albert2006, albert2008}, GeV \citep{abdo2009}, X-ray \citep{paredes1997, leahy2001}, optical H$\alpha$ \citep{grundstrom2007}, and radio \citep{gregory1978, taylor1982}. 

Periodic radio emission from LS I +61 303 was first reported by \citet{taylor1982}.  They defined the arbitrary reference for zero phase at HJD 2,443,366.775 that remains the conventional definition for this system.  The binary orbital period ($P = 26.4960 \pm 0.0028$ d) has been determined from the periodic radio outbursts that peak near $\phi (\rm TG) = 0.6-0.8$ \citep{gregory2002}, and this period is confirmed by optical spectroscopy \citep{aragona2009}.  Periastron occurs at $\phi (\rm TG) = 0.275$ \citep{aragona2009}.  

A number of authors have observed an overall increase in the H$\alpha$ emission line strength and an unusual red shoulder present in that line at the phase of radio maximum \citep{paredes1994, zamanov1999, liu2000, liu2005, grundstrom2007, stoyanov2008}.  \citet{paredes1994} and \citet{liu2000} proposed that an accretion disk around the compact companion may contribute to the H$\alpha$ emission at this phase.  On the other hand, \citet{grundstrom2007} proposed that this periodic emission feature is attributed to a spiral density wave in the Be star disk that is accreted by the compact companion shortly after periastron. There is a correlation between the X-ray flux and the disk emission strength during the orbit (see Figure 5 from \citealt{grundstrom2007}).  The large orbit-to-orbit variations in H$\alpha$ are quite similar to the variability of the radio outbursts. 

 

We recently obtained an extensive collection of red optical spectra of LS I +61 303 to determine an updated orbital ephemeris for the spectroscopic binary \citep{aragona2009}.  These observations, reviewed in \S2, also recorded the evolution of the H$\alpha$ emission during a full orbital cycle.  In \S3 we discuss the short-term (intra-orbit) and long-term variability of the H$\alpha$ emission.  Finally, we discuss the origin of the emission in \S4.


\section{Observations}

During 2008 October and November, we obtained 83 red optical spectra of LS I +61 303 at the KPNO coud\'e feed (CF) telescope over 35 consecutive nights.  These observations are described in detail by \citet{aragona2009}.  To summarize, the spectra covered a wavelength range of 6400--7050 \AA~with a resolving power $R = \lambda/\Delta \lambda \sim 12,000$.   We generally obtained 2--3 spectra of LS I +61 303 each night.  The spectra included only two prominent stellar features, H$\alpha$ and \ion{He}{1} $\lambda6678$, as well as the interstellar line at 6613 \AA.  
The CF spectra were calibrated using standard procedures in IRAF\footnote{IRAF is distributed by the National Optical Astronomy Observatory, which is operated by the Association of Universities for Research in Astronomy (AURA) under cooperative agreement with the National Science Foundation.} and were interpolated onto a common heliocentric wavelength grid.  To rectify the spectra to a unit continuum, we omitted a wide region surrounding the emission line features to avoid any possible contamination of the emission line profile shapes.  

We observed LS I +61 303 again on UT dates 2009 October 17--19, or $0.654 \le \phi (\rm TG) \le 0.733$, at the Wyoming Infrared Observatory (WIRO) using the Long Slit Spectrograph.  We used the LS-1 grating in $2^{nd}$ order, with the KG3 order-sorting filter, and these spectra cover the wavelength range 5400--6800 \AA~with $R \sim 4500$.  Due to variable clouds each night, we used exposure times of 300--900 s to obtain a total of 10 spectra during the run.  We also obtained a CuAr lamp exposure just before each science exposure.  The WIRO spectra were bias corrected, flat fielded, and wavelength calibrated using standard procedures in IRAF.  Finally, we rectified the WIRO spectra to a unit continuum using line-free regions and interpolated them to a common heliocentric wavelength grid.


\section{H$\alpha$ Emission Properties}

\begin{figure}
\includegraphics[angle=0,scale=0.7]{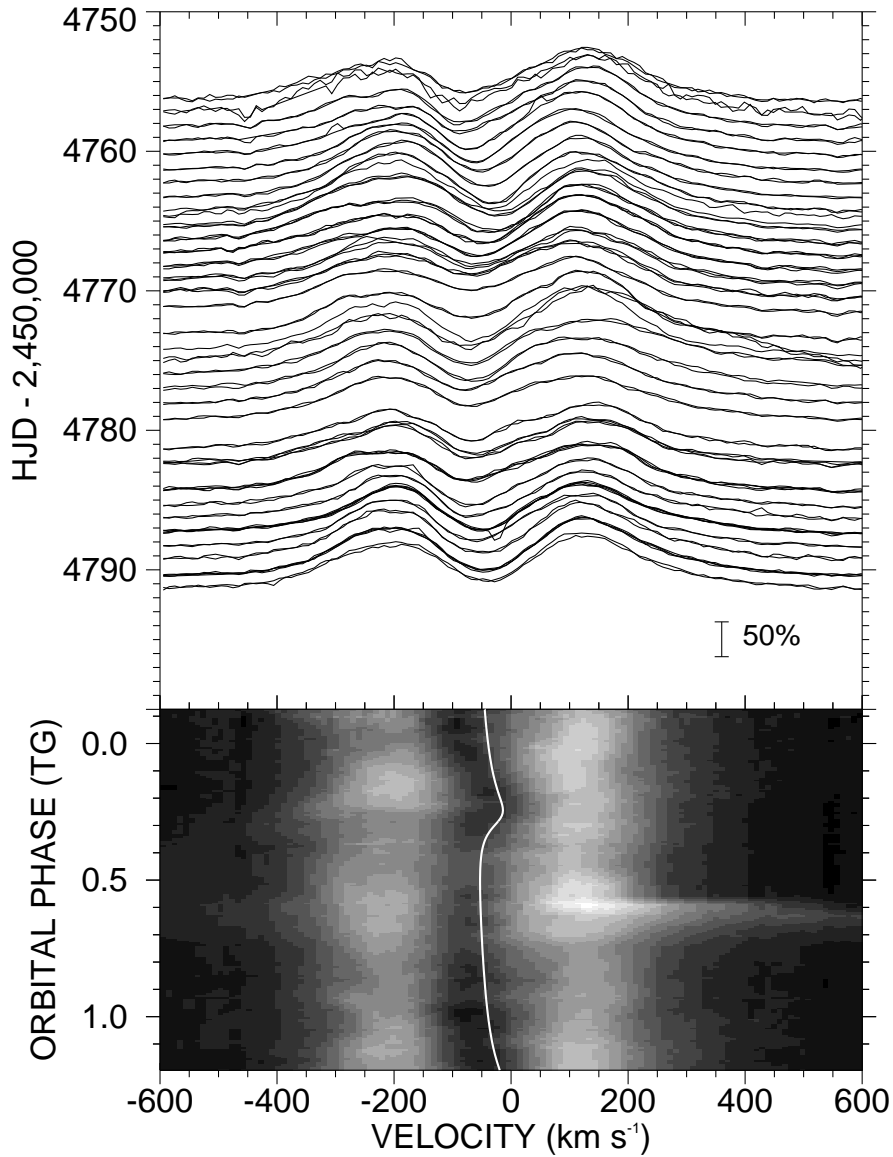} 
\\
\caption{The upper plot shows the H$\alpha$ line profile of LS I +61 303 over our continuous 35 nights of observation, sorted by HJD, and the lower plot shows a gray-scale image of the same line.  Note that the lower plot of the gray-scale spectra are \textit{not} folded by orbital phase but are placed in the same chronological order as the upper plot, with the orbital phases indicated.  The intensity at each velocity in the gray-scale image is assigned one of 16 gray levels based on its value between the minimum (dark) and maximum (bright) observed values.  The intensity between observed spectra is calculated by a linear interpolation between the closest observed phases.  The solid white line shows the theoretical $V_r$ curve solution \citep{aragona2009}.  
\label{hagray} }
\end{figure}

Since the orbital period of LS I +61 303 is only 26.4960 d \citep{gregory2002}, our 2008 collection of CF spectra enable the first detailed study of the H$\alpha$ emission throughout an entire, single orbit.  We show in Figure \ref{hagray} the H$\alpha$ line profiles and a gray-scale image of this line over the span of that observing run.  Some glitches that do not affect the emission line profiles have been removed for clarity.  Note that neither the line profiles nor gray-scale plots are folded by orbital phase, but rather they reveal true chronological variations in the line profile behavior as a function of HJD and the corresponding orbital phase.  

A prominent emission feature stands out as a ``red shoulder'' in the H$\alpha$ line profile between periastron and apastron.  The feature first appears in our spectra from UT date 2008 November 6, at $\phi (\rm TG) = 0.592$.  By the following night, the red component is broader still.  The extension subsides over the next two nights, and vanishes by UT 2008 November 10, $\phi (\rm TG) = 0.740$.    
Although this unusual red shoulder was detected in eight spectra over four nights, we were initially concerned that the spectra might be contaminated by scattered light or some other instrumental effect, or that it could be an artifact of our rectification procedure.  We examined all of the raw CF spectra carefully, including other targets observed on the same nights using the same instrumental configuration, and we repeated our rectification of the complete set of LS I +61 303 spectra to exercise maximum caution in the continuum fitting process.  In fact, this unusual red shoulder is not the result of contamination or a transient optical flare; it is a persistent feature near this orbital phase \citep{paredes1994, liu2000, grundstrom2007}.  

\begin{figure}
\includegraphics[angle=90,scale=0.7]{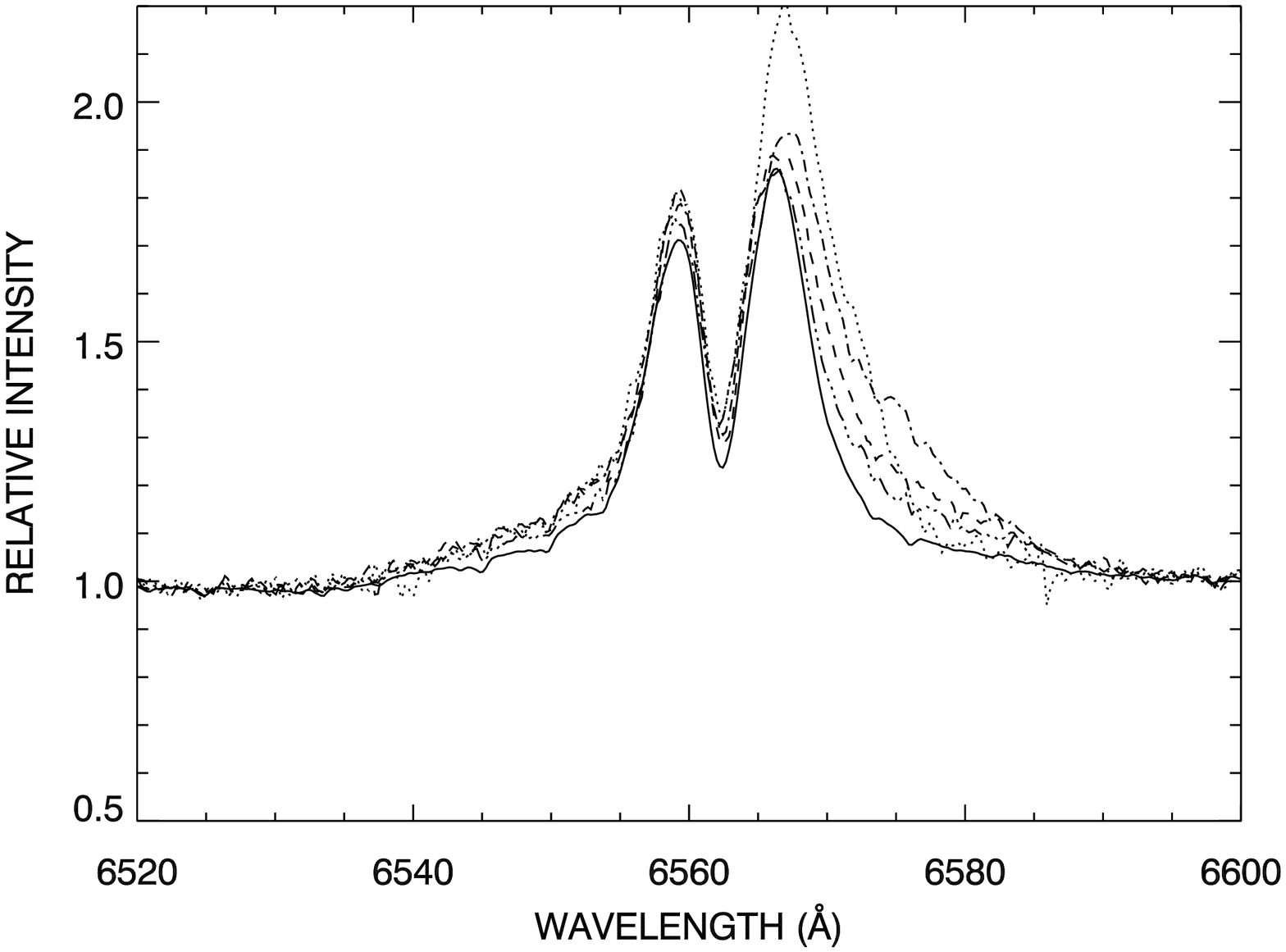} 
\\
\caption{The mean CF H$\alpha$ line profiles from $\phi (\rm TG) = 0.59$ (dotted line), $\phi (\rm TG) = 0.63$ (dot-dashed line), $\phi (\rm TG) = 0.67$ (dashed line), and $\phi (\rm TG) = 0.70$ (dot-dot-dot-dashed line) are compared to the mean CF spectrum (solid line).  All spectra have been shifted to the rest velocity of the optical star.  
\label{halphaCF} }
\end{figure}

\begin{figure}
\includegraphics[angle=90,scale=0.7]{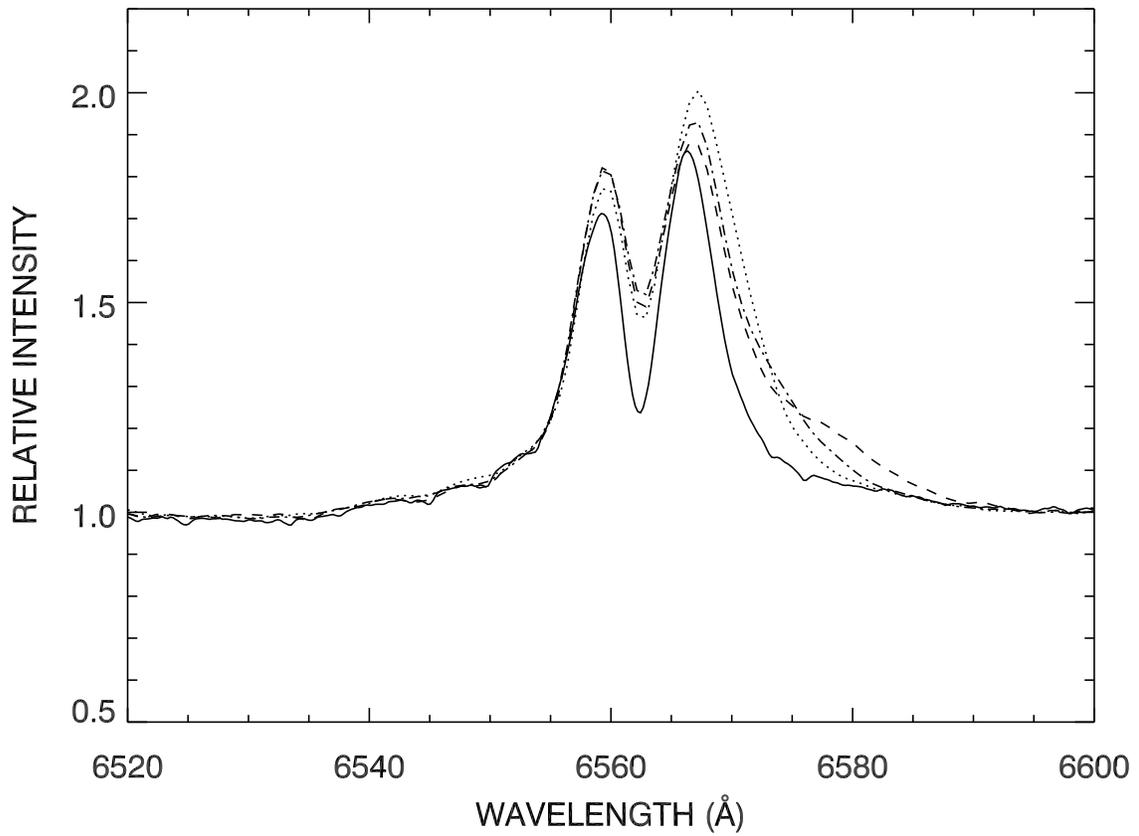} 
\\
\caption{The mean WIRO H$\alpha$ line profiles from $\phi (\rm TG) = 0.66$ (dotted line), $\phi (\rm TG) = 0.69$ (dot-dashed line), and $\phi (\rm TG) = 0.73$ (dashed line) are compared to the mean CF spectrum (solid line).  All spectra have been shifted to the rest velocity of the optical star.  
\label{wiro} }
\end{figure}

The H$\alpha$ line profiles of the WIRO spectra are qualitatively similar to the 2008 CF spectra at comparable orbital phases, shown in Figures \ref{halphaCF} and \ref{wiro}.  A similar, but significantly weaker, extended red-shifted profile is marginally visible near $\phi (\rm TG) \sim 0.6-0.7$ during 1998 August, 1999 October (immediately followed by the nearly complete ionization of the disk), 1999 November, and 2000 December \citep{grundstrom2007}.   It is apparent that over the past decade, the orbit-to-orbit variations in the H$\alpha$ emission line profile are quite large.  

We measured the equivalent width of H$\alpha$, $W_{\rm H\alpha}$, for each spectrum by directly integrating over the line profile.  (We use the convention that $W_{\rm H\alpha}$ is negative for an emission line.)  The errors in $W_{\rm H\alpha}$ are typically about 10\% due to noise and placement of the continuum.  Figure \ref{eqw} shows that during our Coud\'e Feed run, $W_{\rm H\alpha}$ decreases slightly just before periastron.  Since $W_{\rm H\alpha}$ is correlated to the radius of a Be star's circumstellar disk \citep{grundstrom2006}, we interpret the decline in emission as a slight decrease in disk radius as gas is stripped away by the compact companion.  $W_{\rm H\alpha}$ then rises dramatically with the onset of the red shoulder emission component near $\phi (\rm TG) \sim 0.6$.  

Figure \ref{eqw} also compares our recent $W_{\rm H\alpha}$ with those measured by \citet{grundstrom2007}.  Their data were accumulated over six different observing runs over 1998--2000, and the long term differences in emission strength are substantial.  Most of their runs do not provide an opportunity to compare the emission strength at both the periastron and subsequent quadrature phases.  However, their first run over HJD 2451053--2451065 (1998 August) has the best coverage of this orbital range, and the stronger emission strength at periastron suggests that the Be star's disk was slightly larger in radius than in 2008.  The subsequent rise in the 1998 emission occurs sooner than in 2008, but gaps in the 1998 coverage do not allow us to determine whether the emission peak was comparable.  

\citet{stoyanov2008} also obtained 110 $W_{\rm H\alpha}$ measurements over multiple years (see their Figure 2).  Although they do not distinguish between orbital cycles, they find that that the emission can peak anywhere between $0.35 \le \phi(\rm TG) \le 0.7$.  The strong emission episodes never seem to occur past apastron.

Smaller temporal changes in the 2008 CF spectra suggest additional H$\alpha$ emission variability, so we subtracted the mean emission line profile to investigate the residuals carefully.  Although \citet{grundstrom2007} did not observe any radial velocity variations in the peaks or wings of H$\alpha$, we do see them in our dataset, especially in the central trough of the line.  Therefore we formed a mean H$\alpha$ line profile by shifting each spectrum to the star's rest velocity.  Then  we shifted the mean to the orbital velocity of the star according to the ephemeris of \citet{aragona2009} and subtracted it to produce H$\alpha$ emission residuals, or difference spectra, shown in Figure \ref{hadiff}.  We also produced a version of Figure \ref{hadiff} without accounting for orbital motion, and the results were not significantly different.  During about half of the orbit, $0.9 \le \phi (\rm TG) \le 0.6$, the difference spectra reveal a partial S-shaped pattern similar to a spiral density wave that is commonly observed in Be star disks \citep{porter2003}.  \citet{zamanov1996} also observed a strong blue peak near $\phi(\rm TG) = 0.23$, which supports the development of a spiral density wave near periastron.  After this phase, the peculiar red shoulder develops.

The observed behavior of the spiral density wave and the red shoulder emission components are quite different.  To investigate whether these features form in the circumstellar disk of the optical star or in an accretion disk around the compact object, we used Gaussian fits of the peak residual emission in our difference spectra to measure its radial velocity, $V_r$, and full width half maximum, FWHM.  In some cases, the difference spectra did not clearly show a distinctive peak, so we omitted those spectra from measurement.  We also omitted measurements in cases where the residual emission profile deviated substantially from a symmetric, Gaussian profile.  Many of the red shoulder emission profiles were not included as a result.  

Between the orbital phases $0.59 \le \phi (\rm TG) \le 0.63$, our Gaussian fits of the red shoulder measure its $V_r$ between 300--443 km~s$^{-1}$.  The optical star has $V_r \sim -52$ near these phases, and a center of mass velocity $\gamma = -41.41$ km~s$^{-1}$ \citep{aragona2009}.  Assuming a $12.5 M_\odot$ optical star \citep{casares2005a} and a $1.4 M_\odot$ neutron star, the expected radial velocity of the neutron star (and any associated accretion disk) should only be $\sim  54$ km~s$^{-1}$.  If the $V_r$ of the red shoulder was due to orbital motion, this would require a mass ratio $q = M_1/M_2 \gtrsim 25$, which is unlikely.  We conclude that the observed $V_r$ of the red emission shoulder does not correspond to the orbital motion of the compact companion as proposed by \citet{liu2000}.  

The FWHM of the residual emission peaks are shown in Figure \ref{fwhm}.  We estimate that the errors in our FWHM measurements are about 50 km~s$^{-1}$ due to noise intrinsic to the difference spectra and minor deviations from the Gaussian profile used in our fits.  \citet{grundstrom2007} measured a projected rotational velocity $V \sin i = 104 \pm 5$ km~s$^{-1}$ for LS I +61 303.  The true equatorial rotational velocity is $V_{eq}$, and the base of the Keplerian circumstellar disk should rotate at the critical velocity $V_{crit}$, where $V_{eq}/V_{crit} \approx 0.85$ for a typical Be star \citep{mcswain2008}.  Therefore emission that originates in the circumstellar disk of LS I +61 303 should have FWHM $\lesssim 2 V \sin i / (V{eq}/V_{crit}) = 329$.  During orbital phases $0 \le \phi (\rm TG) \le 0.5$ and $0.7 \le \phi (\rm TG) \le 1.0$, the low FWHM is generally consistent with a spiral density wave in the Be star disk.  On the other hand, the FWHM of the red shoulder emission component is significantly higher, suggesting a more turbulent region of gas.  

The $W_{\rm H\alpha}$ and FWHM of the red shoulder emission component suggest that this orbital feature does not originate from within the Be star's disk.  From the $V_r$ of this feature, we also rule out orbital motion of an accretion disk surrounding the compact companion.  In the next section, we discuss other possible origins for this feature.


\section{Discussion}

\citet{paredes1994} first proposed that red shoulder of emission may be due to an accretion disk around the compact companion.  
\citet{romero2007} present a three-dimensional smoothed particle hydrodynamics simulation of accretion onto the neutron star in LS I +61 303.  They predict that spiral structure is induced both in the accretion disk and the Be circumstellar disk near periastron.  In their simulations, the accretion rate peaks with a phase delay of about $0.3 P_{orb}$.  The short duration of the red shoulder emission occurs about 0.3 phase past periastron, consistent with their simulations.  However, H$\alpha$ emission from an accretion disk is not generally visible in the optical spectra of other HMXBs, and the radial velocity of the red shoulder is inconsistent with the orbital motion of an accretion disk, so the accretion disk itself is an unlikely origin for this emission feature.  

Rather than an accretion disk source, \citet{grundstrom2007} propose the development of a tidal stream within the Be star disk near periastron, which extends beyond the truncation radius of the Be star disk and into the vicinity of the compact companion.  Such a tidal stream is consistent with the simulations of \citet{romero2007}.  When the red shoulder appears at  $\phi (\rm TG)$ of 0.59--0.73, we may be looking somewhat down the throat of the resulting mass stream as the compact companion is receding from the observer; the relative orbital geometry is shown in Figure \ref{orbit} using the recent orbital ephemeris by \citet{aragona2009}.  

\begin{figure}
\includegraphics[angle=90,scale=0.7]{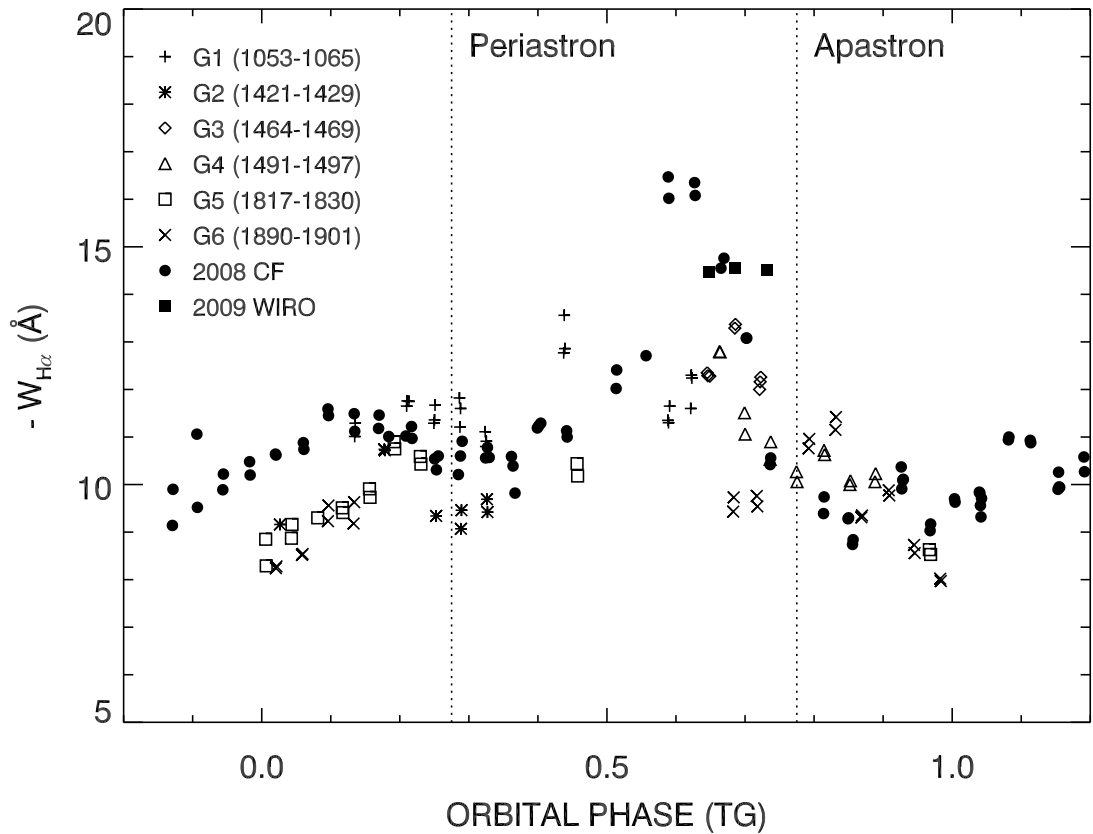} 
\\
\caption{Variations in $W_{\rm H\alpha}$ as a function of orbital phase.  Measurements from the six runs of Grundstrom et al.\ (2007; identified by the date HJD-2,450,000) and the mean $W_{\rm H\alpha}$ from each night of our 2009 WIRO run are shown folded by orbital phase.  The consecutive set of 2008 CF measurements have not been folded but are placed in their true chronological order.  The times of periastron and apastron are marked with vertical dotted lines.  
\label{eqw} }
\end{figure}

\begin{figure}
\includegraphics[angle=0,scale=0.7]{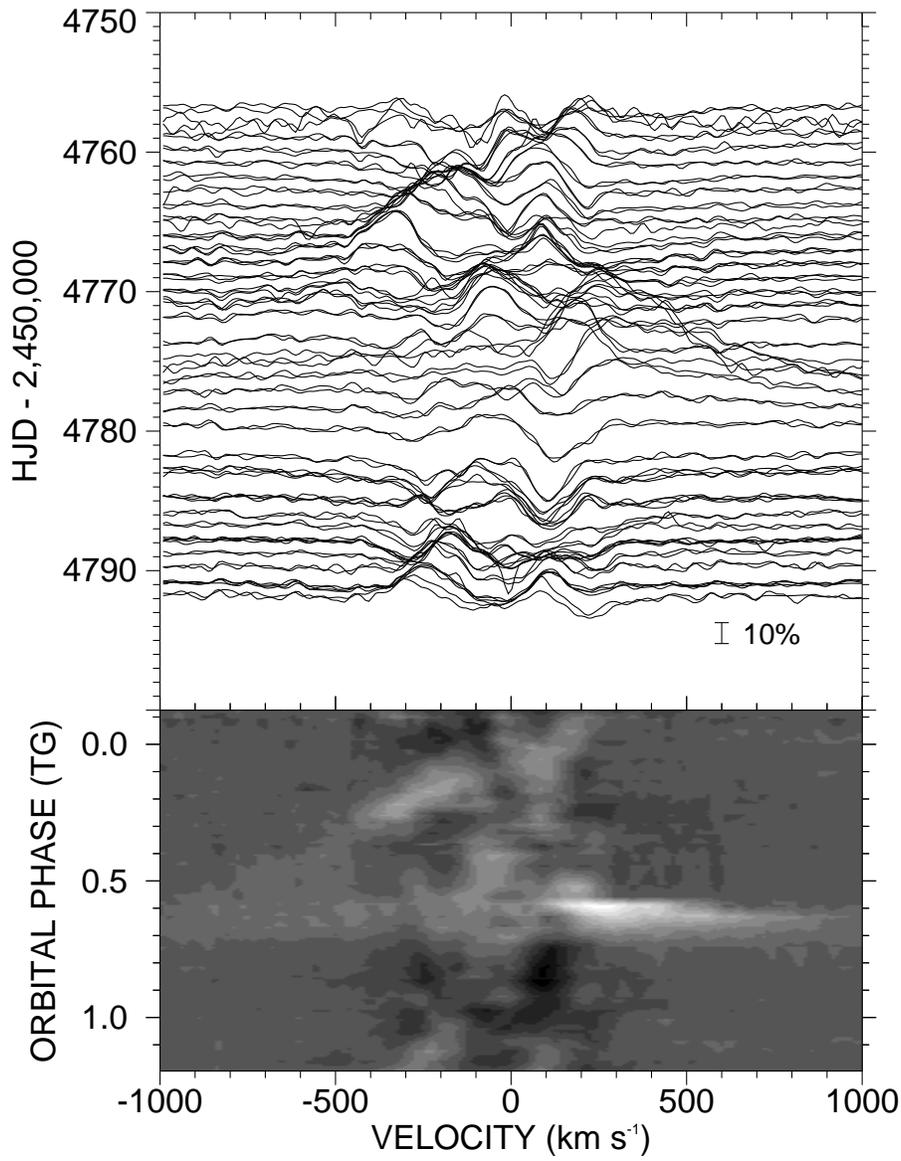} 
\\
\caption{Emission residuals, or difference spectra, in the same format as Figure \ref{hagray}.  The difference spectra have been smoothed for clarity. 
\label{hadiff} }
\end{figure}

\begin{figure}
\includegraphics[angle=90,scale=0.7]{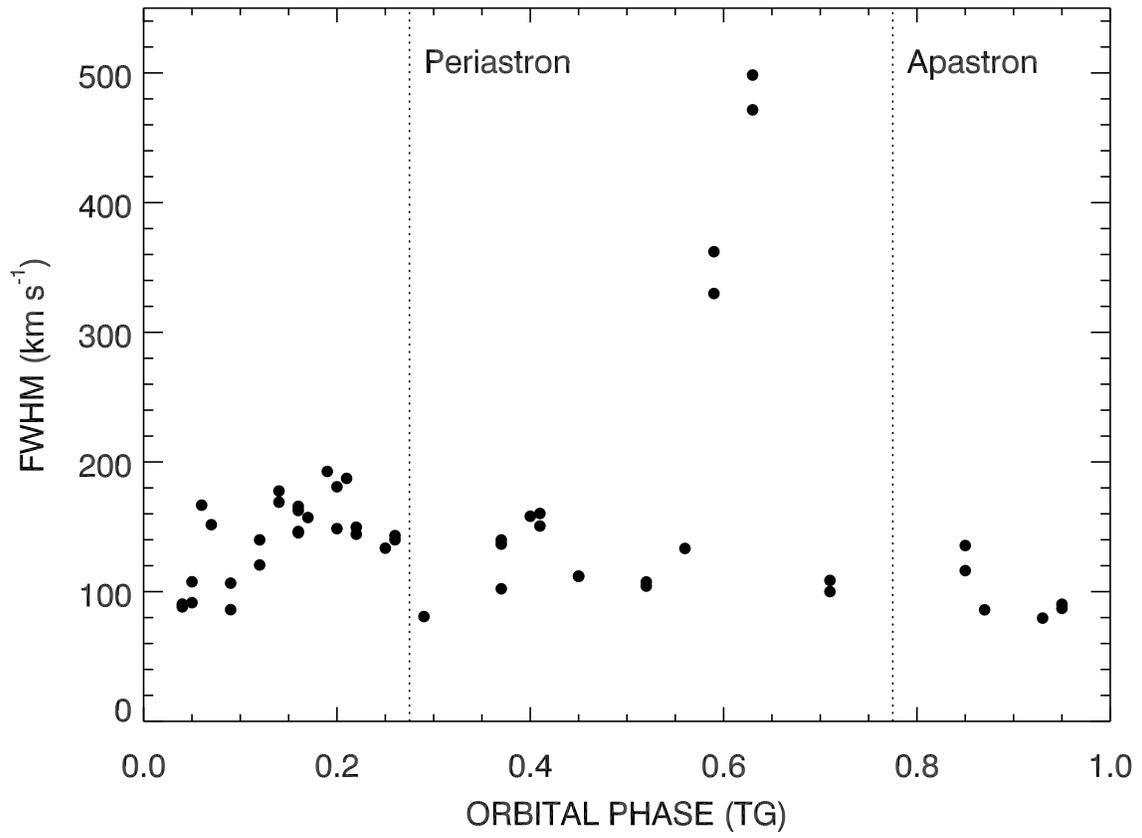} 
\\
\caption{FWHM of the peak residual emission measured from Gaussian fits of the difference spectra.  
\label{fwhm} }
\end{figure}

\begin{figure}
\includegraphics[angle=90,scale=0.8]{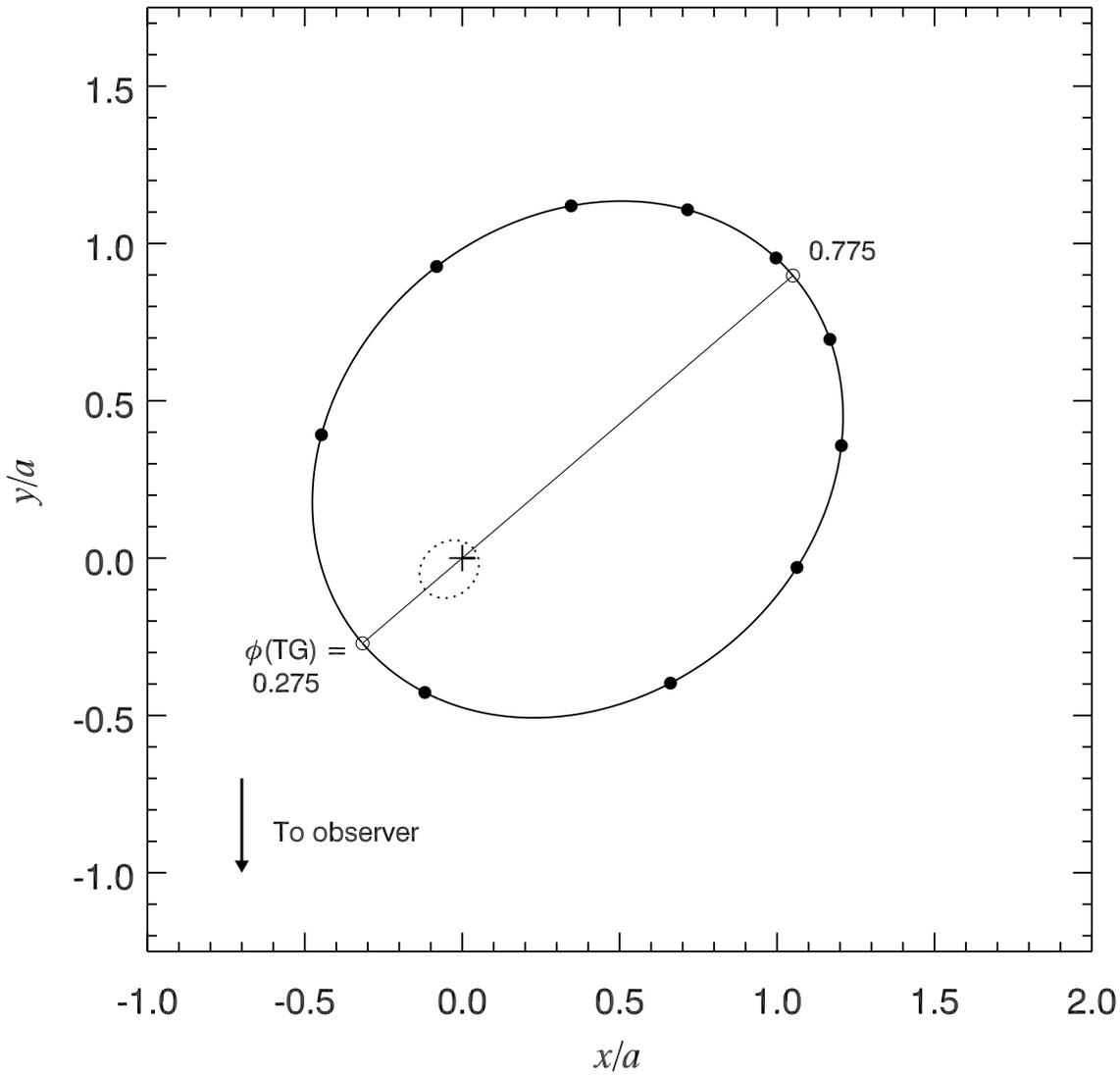} 
\\
\caption{Orbital geometry of LS I +61 303, adapted from \citet{aragona2009}.  The relative orbits ($r/a$) of the $12.5 M_\odot$ optical star (dotted line; \citealt{casares2005a}) and a presumed $1.4 M_\odot$ neutron star companion (solid line) are shown.  The center of mass is indicated with a cross, and the thin solid line is the orbital major axis.  Points around the orbit indicate steps of 0.1 in orbital phase, and the phases of periastron and apastron are labeled.  The stars move counterclockwise in the figure.    
\label{orbit} }
\end{figure}

If the red shoulder of emission is produced by a tidal stream, we should observe a correlation between the disk radius just before periastron (as measured from $W_{\rm H\alpha}$ at this phase) and the emission strength of the mass stream.  This is not observed over multiple epochs of $W_{\rm H\alpha}$ measurements (Fig.\ \ref{eqw}).  Furthermore, the stream should be relatively collimated as it flows across the inner Lagrangian point.  The dynamical trajectory of the stream should not vary from orbit to orbit, so the timing of the emission peak should be constant during each orbit, but there is no evidence of any consistency over many orbits.  While a tube of gas would appear optically thick while it is oriented along our line of sight, the emitting area of the mass stream is unlikely to be large enough to cause up to 50\% increase in the emission strength relative to the normal Be star's disk emission.  Furthermore, the broad FWHM of this feature does not support a collimated mass flow onto the neutron star (Fig.\ \ref{fwhm}).  



The red shoulder emission could be formed as the induced tidal stream falls back onto the Be circumstellar disk.  Since the more massive Be star dominates the gravitational field, most of the tidally disturbed material will not be accreted, instead crashing back onto the Be disk.  Around $\phi (\rm TG) \sim 0.6$, the infalling gas would be moving away from the observer and would impact the redshifted part of the Be disk.  The density would increase quickly as the plume hit the disk and the emission would increase because of the density squared dependence.  If this origin is correct, the orbit-to-orbit variation of the red shoulder behavior may be due to the changes in the Be disk gas density at the azimuth and time near the position of the companion at periastron.  This may be quite variable since the gas involved would be in the tenuous, outer reaches of the disk.  If the red shoulder does come from the outer part of the Be disk facing the companion, the mean densities there may be too low to produce the He I 6678 emission (usually
formed in the inner, high density regions of the disk).  This is probably consistent with the lack of a red shoulder in \ion{He}{1} $\lambda 6678$  (\citealt{aragona2009}; Fig. 7).  

We can estimate the infall velocity of the tidal stream by assuming that its circular velocity is comparable to the periastron velocity of the neutron star, 281 km~s$^{-1}$.  Disk gas would be launched outwards at periastron, and then fall back towards the other side of the disk with a radial velocity of about $+281 - 41 = +240$ km~s$^{-1}$ (correcting for the systemic velocity of the binary;\citealt{aragona2009}).   Our first measurement of the red shoulder is at $V_r = +293$ km~s$^{-1}$ (Table 1),
which is larger, but the disagreement is not too bad considering our simple model.

A fourth option for the origin of the H$\alpha$ shoulder could be a cone-shaped wind shock region.  LS I +61 303 may contain a shrouded pulsar whose relativistic wind interacts with the optical star's wind to form a wind shock region \citep{dubus2006, dubus2010}.  Isolated pulsar wind nebulae occasionally exhibit H$\alpha$ spectra typical of a Balmer-dominated shock (BDS), including the presence of both narrow ($\sim 10$ km~s$^{-1}$) and broad ($\sim 500 - 1000$ km~s$^{-1}$) hydrogen emission line components \citep{heng2010}.  While the blending of the Be star's emission with the red shoulder makes it difficult to identify both H$\alpha$ components of the BDS in LS I +61 303, the unusual broadness of the red shoulder emission is consistent with a BDS.  While many pulsar wind nebulae do not exhibit such a BDS, the high density of the stellar wind surrounding the pulsar wind nebula in LS I +61 303 increases the chances of forming such a shock \citep{heng2010}.  The temporary nature of the red shoulder may suggest that the BDS only forms when the high density tidal stream interacts with the neutron star.  Based on the large, positive $V_r$ of the red shoulder emission observed in this system, the flow of gas along the shock cap should be oriented away from the observer.   However, \citet{romero2007} modeled such a wind shock region in LS I +61 303, but they found difficulties reconciling the observed morphology of the radio emission and the spectral energy distribution with their models.  

\citet{dubus2010} advocate a cometary shock flow in LS I +61 303 trailing the neutron star's orbit (and thus opening toward the observer near $\phi (\rm TG) \sim 0.6$) if the compact object interacts with gas from the Be circumstellar disk that is moving more slowly in a corotating frame.  According to their interpretation, we should observe blue-shifted emission from the gas flow, contradicting the observed red shoulder.  The orbital velocity of the neutron star is given by 
\begin{equation}
V_{orbit} = \sqrt{ \frac{G M_\star^3}{(M_\star + M_{NS})^2}  \left ( \frac{2}{r_{NS}} - \frac{1}{a_{NS}} \right ) }
\end{equation}
where $M_{NS}$ and $M_\star$ are the masses of the neutron star and optical star, $r_{NS}$ is the neutron star's distance from the center of mass, and $a_{NS}$ is the neutron star semimajor axis \citep{hilditch2001}.  Compare this to the approximately circular Keplerian orbit of the circumstellar disk material at the neutron star's location,
\begin{equation}
V_{disk} = \sqrt{  \frac{G M_\star}{r_{NS}} }
\end{equation}
These velocities are equivalent for 
\begin{equation}
r_{NS} = 2 a_{NS} - \left ( \frac{M_\star + M_{NS}}{M_\star} \right )^2 a_{NS}
\end{equation}
Assuming $M_\star = 12.5 \: M_\odot$ \citep{casares2005a} and $M_{NS} = 1.4 \: M_\odot$ with $P = 26.4960$ d, then $a_{NS} = 80.9 \: R_\odot$.  The neutron star would move faster than the circumstellar disk for $0.194 < \phi (TG) < 0.356$, only very close to periastron.  Accounting for only the circular component of the neutron star's orbit (rather than the total orbital speed) changes these results only slightly.  If the Keplerian disk extended out to the neutron star's location, the disk would actually stream past the slow moving neutron star near apastron, potentially reversing the orientation of the shock flow in agreement with our observations of the H$\alpha$ emission.  However, the Be circumstellar disk should be truncated within the periastron distance of the neutron star.  Any stellar wind interacting with the neutron star near $\phi (\rm TG) \sim 0.6$ would be oriented radially away from the Be star, so any comet-shaped interaction region would be skewed away from the line of sight.  


If the red shoulder forms in an optically thin region, we should observe flux from other orbital phases unless the emission is only temporary.  A short-duration red shoulder might be formed by any one of several mechanisms:  the formation of a short-duration emitting structure, like a bullet of gas being ejected (as in SS 433; \citealt{gies2002}), a strong beaming effect that causes the feature to be observed only at particular orientations,  or the complete ionization of the emitting structure at most other orbital phases.  Since we do not observe relativistic velocities, a bullet of ejected gas or beaming effects are unlikely.  Ionization effects would make more sense if the emission was observed more symmetrically about apastron.  In our opinion, the most likely mechanism to form a short-duration emission structure occurs when the induced spiral density wave extends across the full binary separation \citep{romero2007}.  Whether the tidal stream falls back onto the Be circumstellar disk or interacts to form a temporary pulsar wind nebula is somewhat inconclusive.  We recommend further simulations to study the evolution of the tidal stream throughout the entire orbit, and we urge further observations in an effort to conclusively detect a pulsar in this system \citep{rea2010}.


\acknowledgments

We are grateful to the University of Wyoming, especially Chip Kobulnicky and Dan Kiminki, for providing telescope time and observing support at WIRO.   We also thank Di Harmer and the staff at KPNO for their hard work to schedule and support the Coud\'e Feed observations.  Christina Aragona, Tabetha Boyajian, Amber Marsh, and Rachael Roettenbacher helped collect the spectra presented here and should be cheered for their heroic efforts.  We also appreciate helpful conversations with George McCluskey regarding these results.  This work is supported by NASA DPR numbers NNX08AV70G, NNG08E1671, and NNX09AT67G.  MVM is grateful for an institutional grant from Lehigh University.  The research work of DRG is supported by the National Science Foundation
under Grants No.~AST-0506573 and AST-0606861.

{\it Facilities:} \facility{KPNO:CFT ()}, \facility{WIRO ()}


\begin{deluxetable}{lcccc}
\tablewidth{0pt}
\tablecaption{Properties of H$\alpha$ Emission \label{measurements} }
\tablehead{
\colhead{HJD} &
\colhead{Orbital} & 
\colhead{$W_{\rm H\alpha}$} &
\colhead{FWHM of Difference Spectrum} &
\colhead{$V_r$ of Difference Spectrum} \\
\colhead{($-$2,450,000)} &
\colhead{Phase (TG)} &
\colhead{(\AA)}         &
\colhead{(km s$^{-1}$)} 	&
\colhead{(km s$^{-1}$)} }
\startdata
 4756.7256 & 0.87 & \phn $-9.14$ & \phn  86.0  & \phs    223.3  \\
 4756.7482 & 0.87 & \phn $-9.90$ & \nodata     & \nodata        \\
 4757.6644 & 0.91 &     $-11.06$ & \nodata     & \nodata        \\
 4757.6855 & 0.91 & \phn $-9.52$ & \nodata     & \nodata        \\
 4758.6607 & 0.94 & \phn $-9.89$ & \phn  90.3  & \phs    179.9  \\
 4758.6827 & 0.94 &     $-10.22$ & \phn  87.1  & \phs    183.6  \\
 4759.6877 & 0.98 &     $-10.48$ & \nodata     & \nodata        \\
 4759.7088 & 0.98 &     $-10.20$ & \nodata     & \nodata        \\
 4760.6803 & 0.02 &     $-10.64$ & \nodata     & \nodata        \\
 4760.7014 & 0.02 &     $-10.62$ & \nodata     & \nodata        \\
 4761.7508 & 0.06 &     $-10.88$ &      166.7  & \phs\phn 78.0  \\
 4761.7719 & 0.06 &     $-10.74$ &      151.6  & \phs\phn 74.9  \\
 4762.6955 & 0.10 &     $-11.59$ & \nodata     & \nodata        \\
 4762.7166 & 0.10 &     $-11.45$ & \nodata     & \nodata        \\
 4763.7018 & 0.13 &     $-11.49$ &      177.6  &       $-180.5$ \\
 4763.7229 & 0.13 &     $-11.12$ &      169.0  &       $-182.3$ \\
 4764.6373 & 0.17 &     $-11.18$ & \nodata     & \nodata        \\
 4764.6583 & 0.17 &     $-11.46$ &      157.2  &       $-219.3$ \\
 4765.0282 & 0.18 &     $-11.01$ &      192.7  &       $-246.7$ \\
 4765.6837 & 0.21 &     $-11.02$ &      187.3  &       $-281.9$ \\
 4765.9010 & 0.22 &     $-11.22$ &      144.3  &       $-304.3$ \\
 4765.9221 & 0.22 &     $-10.97$ &      149.7  &       $-308.9$ \\
 4766.7893 & 0.25 &     $-10.54$ &      133.5  &       $-339.9$ \\
 4766.8574 & 0.25 &     $-10.31$ &      140.0  &       $-339.0$ \\
 4766.9291 & 0.26 &     $-10.60$ &      143.2  &       $-339.5$ \\
 4767.7007 & 0.28 &     $-10.21$ & \nodata     & \nodata        \\
 4767.7805 & 0.29 &     $-10.60$ & \nodata     & \nodata        \\
 4767.8459 & 0.29 &     $-10.91$ & \phn  80.8  &       $-398.4$ \\
 4768.7456 & 0.32 &     $-10.56$ & \nodata     & \nodata        \\
 4768.8149 & 0.33 &     $-10.78$ & \nodata     & \nodata        \\
 4768.8841 & 0.33 &     $-10.57$ & \nodata     & \nodata        \\
 4769.7368 & 0.36 &     $-10.59$ &      139.9  & \phn   $-69.5$ \\
 4769.7970 & 0.36 &     $-10.39$ &      102.2  & \phn   $-59.5$ \\
 4769.8738 & 0.37 & \phn $-9.82$ &      136.6  & \phn   $-65.8$ \\
 4770.7309 & 0.40 &     $-11.19$ &      158.1  & \phn   $-34.3$ \\
 4770.7911 & 0.40 &     $-11.24$ &      160.3  & \phn   $-36.6$ \\
 4770.8624 & 0.40 &     $-11.29$ &      150.6  & \phn   $-32.1$ \\
 4771.8542 & 0.44 &     $-11.13$ &      150.6  & \phn   $-32.1$ \\
 4771.8753 & 0.44 &     $-11.00$ &      111.9  & \phn   $-45.7$ \\
 4773.7536 & 0.51 &     $-12.02$ &      104.3  & \phs    190.9  \\
 4773.7752 & 0.51 &     $-12.41$ &      107.5  & \phs    193.6  \\
 4774.9024 & 0.56 &     $-12.71$ &      133.3  & \phs    181.7  \\
 4775.7528 & 0.59 &     $-16.47$ &      362.1  & \phs    293.2  \\
 4775.7759 & 0.59 &     $-16.02$ &      329.9  & \phs    295.5  \\
 4776.7648 & 0.63 &     $-16.35$ &      471.6  & \phs    424.8  \\
 4776.7862 & 0.63 &     $-16.08$ &      498.4  & \phs    435.7  \\
 4777.7764 & 0.67 &     $-14.55$ & \nodata     & \nodata        \\
 4777.8919 & 0.67 &     $-14.76$ & \nodata     & \nodata        \\
 4778.7466 & 0.70 &     $-13.08$ &      108.6  & \phn   $-15.1$  \\
 4778.7677 & 0.70 &     $-13.08$ &      100.0  & \phn\phn $ -8.7$ \\
 4779.6636 & 0.74 &     $-10.42$ & \nodata     & \nodata        \\
 4779.6851 & 0.74 &     $-10.56$ & \nodata     & \nodata        \\
 4781.7132 & 0.81 & \phn $-9.39$ & \nodata     & \nodata        \\
 4781.7344 & 0.81 & \phn $-9.74$ & \nodata     & \nodata        \\
 4782.6531 & 0.85 & \phn $-9.28$ &      116.2  & \phn   $-86.4$ \\
 4782.6747 & 0.85 & \phn $-9.30$ &      135.6  & \phn   $-80.5$ \\
 4782.8258 & 0.86 & \phn $-8.74$ & \nodata     & \nodata        \\
 4782.8470 & 0.86 & \phn $-8.84$ & \nodata     & \nodata        \\
 4784.6952 & 0.93 &     $-10.37$ & \phn  79.6  & \phn   $-17.0$ \\
 4784.7163 & 0.93 & \phn $-9.91$ & \nodata     & \nodata        \\
 4784.7536 & 0.93 &     $-10.09$ & \nodata     & \nodata        \\
 4784.7747 & 0.93 &     $-10.11$ & \nodata     & \nodata        \\
 4785.8022 & 0.97 & \phn $-9.03$ & \nodata     & \nodata        \\
 4785.8233 & 0.97 & \phn $-9.17$ & \nodata     & \nodata        \\
 4786.7401 & 0.00 & \phn $-9.70$ & \nodata     & \nodata        \\
 4786.7612 & 0.00 & \phn $-9.63$ & \nodata     & \nodata        \\
 4787.6918 & 0.04 & \phn $-9.84$ & \phn  90.4  &       $-166.8$ \\
 4787.7129 & 0.04 & \phn $-9.82$ & \phn  88.3  &       $-163.2$ \\
 4787.7340 & 0.04 & \phn $-9.56$ & \nodata     & \nodata        \\
 4787.7551 & 0.04 & \phn $-9.32$ &      107.6  &       $-169.1$ \\
 4787.7762 & 0.04 & \phn $-9.71$ & \phn  91.5  &       $-175.5$ \\
 4788.8069 & 0.08 &     $-10.94$ &      106.6  &       $-200.1$ \\
 4788.8280 & 0.08 &     $-11.00$ & \phn  86.1  &       $-204.3$ \\
 4789.6443 & 0.11 &     $-10.93$ &      139.9  &       $-215.7$ \\
 4789.6654 & 0.11 &     $-10.88$ &      120.6  &       $-212.5$ \\
 4790.7113 & 0.15 & \phn $-9.90$ &      146.4  &       $-238.1$ \\
 4790.7324 & 0.15 &     $-10.26$ &      145.3  &       $-239.0$ \\
 4790.7536 & 0.15 & \phn $-9.95$ &      162.6  &       $-227.1$ \\
 4790.7747 & 0.16 & \phn $-9.94$ &      165.8  &       $-238.1$ \\
 4791.7059 & 0.19 &     $-10.58$ &      180.9  &       $-276.9$ \\
 4791.7270 & 0.19 &     $-10.27$ &      148.6  &       $-286.5$ \\
 5121.8113 & 0.65 &     $-$14.80 & \nodata     & \nodata        \\
 5121.8591 & 0.66 &     $-$14.27 & \nodata     & \nodata        \\
 5121.8812 & 0.66 &     $-$14.35 & \nodata     & \nodata        \\
 5122.8172 & 0.69 &     $-$14.60 & \nodata     & \nodata        \\
 5122.8282 & 0.69 &     $-$14.64 & \nodata     & \nodata        \\
 5122.8562 & 0.69 &     $-$14.43 & \nodata     & \nodata        \\
 5122.8644 & 0.69 &     $-$14.60 & \nodata     & \nodata        \\
 5122.8900 & 0.69 &     $-$14.46 & \nodata     & \nodata        \\
 5122.8978 & 0.70 &     $-$14.45 & \nodata     & \nodata        \\
 5123.9006 & 0.73 &     $-$14.51 & \nodata     & \nodata        \\

\enddata

\end{deluxetable}

\clearpage

\clearpage

\clearpage

\clearpage

\clearpage

\clearpage

\clearpage


\end{document}